\documentclass[a4paper,12pt]{article}
\usepackage[T1]{fontenc}
\usepackage[utf8]{inputenc}
\usepackage{lmodern}
\usepackage{amsthm}
\usepackage{amsmath}
\usepackage{bbold}
\usepackage[dvips]{graphicx}

\title{The Axigluon, a Four-Site Model and the Top Quark Forward-Backward Asymmetry
 at the Tevatron} 
\author{Alfonso R. Zerwekh\\
Instituto de F\'isica, Facultad de Ciencias,\\
  Universidad Austral de Chile\\
 and \\
Centro Cient\'ifico-Tecnol\'ogico de Valpara\'iso\\ 
  Casilla 567, Valdivia, Chile}
\date{}
\begin{document}

\maketitle

\begin{abstract}
Very recently, it has been shown by other authors that the CDF
measurement of the top quark forward-backward asymmetry can be
explained by means of a heavy and broad axigluon. In order to work,
this mechanism needs that the axigluon coupling to the top
quark must be different than the coupling to light quarks and both
must be stronger than the one predicted in classical axigluon
models. In this paper, we argue that this kind of axigluon can be
accommodated in an extended chiral color model we proposed
previously. Additionally, we show that the desired features can be
derived from a simple four-site model with delocalized fermions.  

\end{abstract}

\section{Introduction}

Although the High Energy Physics community is unanimous in considering
the Standard Model (SM) to be incomplete for many and well known
reasons, such as the Higgs mass instability under quantum corrections
or the unexplained hierarchy observed in the masses of the fermionic
sector; it is 
impossible to deny its enormous (and, to some extent, 
unexpected) experimental success.  Indeed, after years of scrutiny and
new physics search, first at the LEP and the Tevatron and now at the
LHC, the SM has not been still seriously challenged by
collider experiments. The only exception today may be the $3.2\sigma$
deviation recently reported by CDF in the top-quark forward-backward
asymmetry ($A^{t\bar{t}}_{FB}$)\cite{Aaltonen:2011kc} (see also
\cite{Aaltonen:2008hc} and \cite{:2007qb}). The fact that the process under
question involves the top quark makes it even more interesting since,
due to its mass which is near the electroweak scale, it has
been suspected for a long time that the top-quark plays a special
r\^ole in nature.

A natural candidate to explain a deviation in $A^{t\bar{t}}_{FB}$ is
an axigluon (other alternatives beyond the SM have been considered in
\cite{Dorsner:2009mq},\cite{Cao:2010zb},\cite{Berger:2011ua},
\cite{Bhattacherjee:2011nr},
\cite{Patel:2011eh},\cite{Grinstein:2011yv},\cite{Barger:2011ih}
and \cite{Gresham:2011dg}). In fact, this is an expected axigluon
signal \cite{Antunano:2007da}. It is true
that it has been argued that the present signal cannot be explained by
an axigluon due to other experimental constraints over
it \cite{Chivukula:2010fk}. Nevertheless, Bai, Hewett, Kaplan and Rizzo have argued that a
heavy and broad axigluon can be a viable explanation of the
$A^{t\bar{t}}_{FB}$ while remaining invisible in the dijet spectrum \cite{Bai:2011ed}.

In their paper, they parameterize the axigluon interaction as
$$
g_s g_A\bar{\psi}\frac{\lambda^a}{2}\gamma_{\mu}\gamma_5 A^{\mu}\psi
$$
where $g_s$ is the usual QCD coupling constant a $g_A$ measures the
deviation of the axigluon coupling constant from QCD. In order to
evade experimental constraint it proves useful to consider that $g_A$
may take different values for light quarks ($g^q_A$) and for the top
quark ($g^t_A$). The analysis presented in \cite{Bai:2011ed} shows that in
order to have a broad axigluon it must occur  that $|g^q_A|,|g^t_A| >1$.
Additionally, in order to escape detection in contact interaction
searches it must happen that $|g^q_A|< |g^t_A|$ (this kind of
non-universality has been also reported in \cite{Ferrario:2009bz}
).  Finally, $g^q_A
g^t_A$ must be negative in order to reproduce correctly the sign of
$A^{t\bar{t}}_{FB}$.

The construction of a model that can incorporate all these conditions,
specially the different coupling constants for different generations,
has proven to be difficult. As a solution, the authors of \cite{Bai:2011ed}
propose a two-site model but they have to include exotic vector-like
quarks in order to satisfy the  $|g^q_A|< |g^t_A|$ constrain.

In this paper, we address the construction of a viable axigluon model
not by adding new fermions but by extending the gauge sector of the
model. In section \ref{sec:effmodel} we show that an effective model,
based on $SU(3)_L \times SU(3)_R$ but with four octet fields
transforming as gauge bosons, has enough structure to accommodate all
the desire features of the axigluon. In section  \ref{sec:4site}, we
show that the essential characteristics of the effective model can be
obtained from a four-site model with delocalized fermions. Finally, in
section \ref{sec:summary}, we summarize our conclusions.

\section{An Effective Model}
\label{sec:effmodel}
We start by recalling an effective extended chiral model we have previously
proposed\cite{Zerwekh:2009vi}. In this
model, we consider an usual $SU(3)_L\times SU(3)_R$ with two pairs of
vectors fields: $l_{\mu}$ and $L_{\mu}$ transforming as gauge fields
of $SU(3)_L$, and $r_{\mu}$ and $R_{\mu}$ transforming as gauge fields
of $SU(3)_R$. The gauge sector of the model (including a non-linear
sigma model sector which is introduced to produce the breaking of
$SU(3)_L\times SU(3)_R$ to $SU(3)_c$) is described by the following
Lagrangian: 

\begin{eqnarray}
\mathcal{L} & = & -\frac{1}{2}\mathrm{tr}\left\{
  G_{L\mu\nu}G_{L}^{\mu\nu}\right\} -\frac{1}{2}\mathrm{tr}\left\{
  G_{R\mu\nu}G_{R}^{\mu\nu}\right\} \notag\\ 
 &  & -\frac{1}{2}\mathrm{tr}\left\{
   \rho_{L\mu\nu}\rho_{L}^{\mu\nu}\right\}
 -\frac{1}{2}\mathrm{tr}\left\{
   \rho_{R\mu\nu}\rho_{R}^{\mu\nu}\right\} \notag\\ 
 &  & +\frac{M^{2}}{g'^{2}}\mathrm{tr}\left\{
   \left(gl_{\mu}-g'L_{\mu}\right)^{2}\right\}
 +\frac{M^{2}}{g'^{2}}\mathrm{tr}\left\{
   \left(gr_{\mu}-g'R_{\mu}\right)^{2}\right\} \notag\\ 
 &  & +\frac{f^{2}}{2}\mathrm{tr}\left\{
   D_{\mu}U^{\dagger}D^{\mu}U\right\} \label{eq:LagMyModel}
\end{eqnarray}
where

\begin{eqnarray*}
G_{L\mu\nu} & = & \partial_{\mu}l_{\nu}-\partial_{\nu}l_{\mu}-ig_{L}\left[l_{\mu},\: l_{\nu}\right]\\
G_{R\mu\nu} & = & \partial_{\mu}r_{\nu}-\partial_{\nu}r_{\mu}-ig_{R}\left[r_{\mu},\: r_{\nu}\right]\\
\rho_{L\mu\nu} & =
& \partial_{\mu}L_{\nu}-\partial_{\nu}L_{\mu}-ig'_{L}\left[L_{\mu},\:
  L_{\nu}\right]\\ 
\rho_{R\mu\nu} & =
& \partial_{\mu}R_{\nu}-\partial_{\nu}R_{\mu}-ig'_{R}\left[R_{\mu},\:
  R_{\nu}\right]\\ 
D_{\mu}U & = & \partial_{\mu}U-ig_{L}l_{\mu}U+ig_{R}Ur_{\mu}
\end{eqnarray*}

In the unitary gauge a non-diagonal mass matrix is explicitly
generated. Although the resulting mass matrix can be exactly diagonalized,
in order to obtain easily manipulable expressions we will assume that
$g'\gg g$ and we will write our results in the first order in $g/g'$.
The eigenvalues, that is, the masses of the physical states are:

\begin{eqnarray*}
m_{G} & = & 0\\
m_{A} & = & \frac{gf}{\sqrt{2}}\\
m_{G'} & = & M\\
m_{A'} & = & M
\end{eqnarray*}

Thus, the physical spectrum is composed of a (exactly) massless gluon,
an axigluon and two degenerate (at this level of approximation) heavy gluon
and axigluon. The normalized mass eigenvectors can be written as:

\begin{eqnarray}
G_{\mu} & = &
\frac{1}{\sqrt{2}}l_{\mu}+\frac{1}{\sqrt{2}}r_{\mu}+\frac{g}{\sqrt{2}g'}L_{\mu}+\frac{g}{\sqrt{2}g'}R_{\mu}\notag\\  
A_{\mu} & = & -\frac{1}{\sqrt{2}}l_{\mu}+\frac{1}{\sqrt{2}}r_{\mu}-\frac{g}{\sqrt{2}g'}\left(1-\frac{m_{A}^{2}}{M^{2}}\right)L_{\mu}+\frac{g}{\sqrt{2}g'}\left(1-\frac{m_{A}^{2}}{M^{2}}\right)R_{\mu}\notag\\
G'_{\mu} & = & \frac{g}{\sqrt{2}g'}l_{\mu}+\frac{g}{\sqrt{2}g'}r_{\mu}-\frac{1}{\sqrt{2}}L_{\mu}-\frac{1}{\sqrt{2}}R_{\mu}\notag\\
A'_{\mu} & = &
-\frac{g}{\sqrt{2}g'}\left(1-\frac{m_{A}^{2}}{M^{2}}\right)^{-1}l_{\mu}+\frac{g}{\sqrt{2}g'}\left(1-\frac{m_{A}^{2}}{M^{2}}\right)^{-1}r_{\mu}+\notag\\
& &+\frac{1}{\sqrt{2}}L_{\mu}-\frac{1}{\sqrt{2}}R_{\mu} 
\label{eq:estadosfisicos}
\end{eqnarray}

Because we have now two fields that transform as gauge fields for
each group ($l_{\mu}$ and $L_{\mu}$ for $SU(3)_{L}$ and $r_{\mu}$
and $R_{\mu}$ for $SU(3)_{R}$) any combination of the form
$g(1-k)l_{\mu}+g'kL_{\mu}$ 
and $g(1-k)r_{\mu}+g'k R_{\mu}$, where $k$ is a new arbitrary
parameter. In principle, it can take different values in the right and
left sectors but, for simplicity we will assume it is the same in both
cases. With this structure it is possible to construct generalized
covariant 
derivatives. This means 
that the Lagrangian describing the gauge interaction of quarks can
be written as:

\begin{eqnarray}
\mathcal{L} & = &
\frac{1}{2}g(1-k)\bar{\psi}l_{\mu}\gamma^{\mu}(1-\gamma_{5})\psi+\frac{1}{2}g'k\bar{\psi}L_{\mu}
\gamma^{\mu}(1-\gamma_{5})\psi+\nonumber  
\\ 
 &  &
 +\frac{1}{2}g(1-k)\bar{\psi}r_{\mu}\gamma^{\mu}(1+\gamma_{5})\psi+\frac{1}{2}g'k\bar{\psi}R_{\mu}
\gamma^{\mu}(1+\gamma_{5})\psi 
 \label{eq:MyLagFerm}
 \end{eqnarray}
Using Lagrangian (\ref{eq:MyLagFerm})
and the definition of the physical fields, we can obtain the terms
of the Lagrangian that couple the gluon and the axigluon to quarks 

\begin{equation}
\mathcal{L}=g_s\bar{\psi}G_{\mu}\gamma^{\mu}\psi+
g_s g_A \bar{\psi}A_{\mu}\gamma^{\mu}\gamma_{5}\psi
\end{equation} 
where:

\begin{equation}
  g_s \equiv \frac{g}{\sqrt{2}}
\end{equation}
and
\begin{equation}
  g_A \equiv 1-\frac{m^2_A}{M^2}k
\end{equation}
Interestingly, the mechanism described just above is related
with fermion delocalization in deconstruction theory \cite{Chivukula:2005bn,Chivukula:2005xm}
 . 

Now we have an axigluon with a modified coupling to quarks. Nothing in
the model prevents that the value of $g_A$ varies from one generation
to another. This implies that in this framework we can easily satisfy
the necessary conditions (in particular,  $|g^q_A|< |g^t_A|$
constrain)  for making the axigluon compatible with existing data and,
at the same time, explain the $A^{t\bar{t}}_{FB}$ data from CDF.     

The key features of the model described in this section are the
presence of two pairs of vector bosons which transform like gauge
fields and delocalized fermions. These properties can be naturally
implemented in a four-site model. The construction of such a model is
the subject of next section.

\section{A Simple Four-Site Model}
\label{sec:4site}
\begin{figure}
\centering
\includegraphics[scale=0.5]{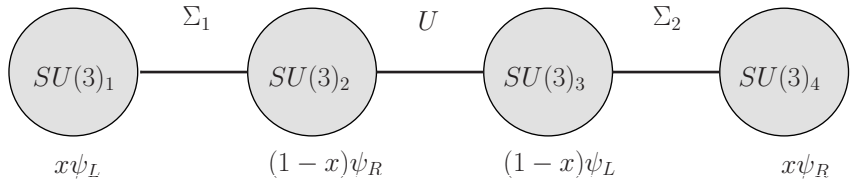}
\caption{Group structure of our simple four-site model}
\label{fig:group}
\end{figure}

Our simple four site model is based on the gauge group $SU(3)_1\times
SU(3)_2 \times SU(3)_3 \times SU(3)_4$ and three link fields
($\Sigma_1$, $U$ and $\Sigma_2$) as shown in figure
\ref{fig:group}. We assume that the scalar fields develop vacuum
expectation values ( $u=\langle \Sigma_1\rangle =\langle
\Sigma_2\rangle$ and $v=\langle U \rangle $ ) breaking down the gauge
symmetry to $SU(3)_c$ which we identify with the usual color group.  

The Lagrangian for the bosonic sector can be written as:

\begin{eqnarray}
\mathcal{L}&=&-\frac{1}{2}\sum_{n=1}^4 \mathrm{tr}\{ F_{n\mu\nu}F_n^{\mu\nu}\}+
\frac{u^2}{2}\mathrm{tr}\{D_{\mu}\Sigma_1^{\dagger}D^{\mu}\Sigma_1 \} + \nonumber\\
& & + \frac{v^2}{2}\mathrm{tr}\{D_{\mu}U^{\dagger}D^{\mu}U \} +
\frac{u^2}{2}\mathrm{tr}\{D_{\mu}\Sigma_2^{\dagger}D^{\mu}\Sigma_2 \}
\end{eqnarray}
where

\begin{eqnarray*}
F_{n\mu\nu}&=&\partial_{\mu}A_{n\nu}-\partial_{\nu}A_{n\mu}-ig[A_{n\mu},A_{n\nu}]\\
D^{\mu}\Sigma_1 &=& \partial_{\mu}\Sigma_1 -igA_{1\mu}\Sigma_1 +ig\Sigma_1 A_{2\mu} \\
D^{\mu}U &=& \partial_{\mu}U -igA_{2\mu}U +igUA_{3\mu} \\
D^{\mu}\Sigma_2 &=& \partial_{\mu}\Sigma_2 -igA_{3\mu}\Sigma_2 +ig\Sigma_2 A_{4\mu}
\end{eqnarray*}
Notice that in order to simplify our equations we have assume that all
groups share the same coupling constant $g$.

As usual, in the unitary gauge ($\Sigma_1=U=\Sigma_2=\mathbb{1}$) a
non-diagonal mass matrix for the gauges bosons explicitly
appears. This mass matrix can be exactly diagonalized, nevertheless,
in benefice of simplicity, we work in the limit where $v\ll u$. In
this limit, the physical spectrum is composed of a (exactly) massless
gluon $G$, a light (in the sense that its mass is proportional to $v$)
axigluon $A$ and two degenerate heavy states ($V_1$ and $V_2$) with
masses proportional to $u$. The original gauge fields can be written
in terms of the  physical fields as follows:  

\begin{eqnarray}
A_{1\mu}=\frac{1}{2}G_{\mu}+\frac{1}{2}A_{\mu}+\frac{1}{2}V_{1\mu}+\frac{1}{2}V_{2\mu}\\
A_{2\mu}=\frac{1}{2}G_{\mu}+\frac{1}{2}A_{\mu}-\frac{1}{2}V_{1\mu}-\frac{1}{2}V_{2\mu}\\
A_{3\mu}=\frac{1}{2}G_{\mu}-\frac{1}{2}A_{\mu}-\frac{1}{2}V_{1\mu}+\frac{1}{2}V_{2\mu}\\
A_{4\mu}=\frac{1}{2}G_{\mu}-\frac{1}{2}A_{\mu}+\frac{1}{2}V_{1\mu}-\frac{1}{2}V_{2\mu}
\end{eqnarray}    

Now we turn our attention to quarks. A crucial feature of our effective
model described in the previous section was the fact that the fermions
coupled to a certain linear combination of gauge bosons. In the
language of deconstruction theory, it correspond to "fermion
delocalization".  The main idea of fermion delocalization
\cite{Chivukula:2005bn,Chivukula:2005xm} is that the
low energy properties of the fermions (for example the color
interaction of quarks) originate from the contribution of many sites
of the underlying gauge group. In other words, the current couples to
a linear combination of gauge fields $\sum_n x_n A_{n\mu}$ with the
constrain $\sum_n x_n=1$ in order to preserve gauge invariance under
the whole group. As shown in figure \ref{fig:group}, we decide to
couple left handed quarks to $A_{1\mu}$ and $A_{3\mu}$ (with weights
$x$ and $1-x$ respectively) and right handed quarks to $A_{2\mu}$ and
$A_{4\mu}$ (with weights $1-x$ and $x$ respectively). In this way, the
interaction Lagrangian for quarks can be written as: 

\begin{eqnarray}
\mathcal{L}_{int}&=& gJ_{L\mu}^a\left[xA^{a\mu}_1+(1-x)A^{a\mu}_3\right]+\nonumber\\
& &  +  gJ_{R\mu}^a\left[xA^{a\mu}_4+(1-x)A^{a\mu}_2\right]
\end{eqnarray}
where $J_{L\mu}^a$ and 
$J_{R\mu}^a$ are the quark left-handed and right-handed currents,
respectively. Notice that we have chosen, for 
simplicity, the
same delocalization parameter $x$ for right handed and left handed
quarks. In terms of the physical fields the interaction Lagrangian
can be written as: 

\begin{eqnarray}
  \label{eq:Lint}
  \mathcal{L}_{int}&=&g_s\bar{\psi}\frac{\lambda^a}{2}\gamma_{\mu}\psi
  G^{a\mu}+
  g_s(1-2x)\bar{\psi}\frac{\lambda^a}{2}\gamma_{\mu}\gamma_5\psi
  A^{a\mu}\nonumber \\
&
&+g_s(2x-1)\bar{\psi}\frac{\lambda^a}{2}\gamma_{\mu}\psi V_1^{a\mu}
-g_s\bar{\psi}\frac{\lambda^a}{2}\gamma_{\mu}\gamma_5\psi
  V_2^{a\mu} 
\end{eqnarray}
where $g_s=g/2$ is the usual strong interaction coupling constant. Our
delocalization pattern has provoked the appearance of a axigluon
interaction which depends on the delocalization parameter $x$. As in
our effective model, nothing prevents that different quark generations
can be differently delocalized. In this way, we can select appropriated
values of $x$ for the top quark and for the light quarks in order to
consistently explain the $A^{t\bar{t}}_{FB}$ measurements.

\section{Summary}
\label{sec:summary}
In summary, we have shown that an extended chiral color model with
extra color octet spin-1 fields has enough structure to describe an
axigluon with appropriate coupling to quarks in such a way that it can
be consistent with dijet and contact interactions searches and, at the
same time, explain the $A^{t\bar{t}}_{FB}$ anomaly reported by
CDF. The fact that this effective model can be re-obtained from a
four-site model, as we have shown, is interesting from the theoretical
point of view since, in principle, it may be related, via dimensional
deconstruction, to five dimensional models.

\section{Acknowledgments}

This work has been partially supported by Fondecyt grant 1070880, the
 Conicyt grant ``Southern Theoretical Physics Laboratory'' ACT-91,
 the Conicyt grant `` Instituto para Estudios Avanzados en Ciencia
y Tecnología'' ACT-119 and Fondecyt grant 1110167. TGD


\begin{thebibliography}{99}
\bibitem{Aaltonen:2011kc}
  T.~Aaltonen {\it et al.} [ CDF Collaboration ],
  ``Evidence for a Mass Dependent Forward-Backward Asymmetry in Top Quark Pair Production,''
  Submitted to: Phys.Rev.D.
  [arXiv:1101.0034 [hep-ex]].

\bibitem{Aaltonen:2008hc}
  T.~Aaltonen {\it et al.} [ CDF Collaboration ],
  ``Forward-Backward Asymmetry in Top Quark Production in ppbar
  Collisions at sqrt{s}=1.96 TeV,'' 
  Phys.\ Rev.\ Lett.\  {\bf 101 } (2008)  202001.
  [arXiv:0806.2472 [hep-ex]].

\bibitem{:2007qb}
  V.~M.~Abazov {\it et al.} [ D0 Collaboration ],
  ``First measurement of the forward-backward charge asymmetry in top quark pair production,''
  Phys.\ Rev.\ Lett.\  {\bf 100 } (2008)  142002.
  [arXiv:0712.0851 [hep-ex]].

\bibitem{Dorsner:2009mq}
  I.~Dorsner, S.~Fajfer, J.~F.~Kamenik, N.~Kosnik,
  ``Light colored scalars from grand unification and the
  forward-backward asymmetry in t t-bar production,'' 
  Phys.\ Rev.\  {\bf D81 } (2010)  055009.
  [arXiv:0912.0972 [hep-ph]].

\bibitem{Cao:2010zb}
  Q.~-H.~Cao, D.~McKeen, J.~L.~Rosner, G.~Shaughnessy, C.~E.~M.~Wagner,
  ``Forward-Backward Asymmetry of Top Quark Pair Production,''
  Phys.\ Rev.\  {\bf D81 } (2010)  114004.
  [arXiv:1003.3461 [hep-ph]].

\bibitem{Berger:2011ua}
  E.~L.~Berger, Q.~-H.~Cao, C.~-R.~Chen, C.~S.~Li, H.~Zhang,
  ``Top Quark Forward-Backward Asymmetry and Same-Sign Top Quark Pairs,''
  
  [arXiv:1101.5625 [hep-ph]].

\bibitem{Bhattacherjee:2011nr}
  B.~Bhattacherjee, S.~S.~Biswal, D.~Ghosh,
  
  [arXiv:1102.0545 [hep-ph]].



\bibitem{Patel:2011eh}
  K.~M.~Patel, P.~Sharma,
  ``Forward-backward asymmetry in top quark production from light
  colored scalars in SO(10) model,'' 
  [arXiv:1102.4736 [hep-ph]].

\bibitem{Grinstein:2011yv}
  B.~Grinstein, A.~L.~Kagan, M.~Trott, J.~Zupan,
  ``Forward-backward asymmetry in t anti-t production from flavour symmetries,''
    [arXiv:1102.3374 [hep-ph]].

\bibitem{Barger:2011ih}
  V.~Barger, W.~-Y.~Keung, C.~-T.~Yu,
  ``Tevatron Asymmetry of Tops in a W',Z' Model,''
    [arXiv:1102.0279 [hep-ph]].

\bibitem{Gresham:2011dg}
  M.~I.~Gresham, I.~-W.~Kim, K.~M.~Zurek,
  ``Searching for Top Flavor Violating Resonances,''
  [arXiv:1102.0018 [hep-ph]].

\bibitem{Antunano:2007da}
  O.~Antunano, J.~H.~Kuhn, G.~Rodrigo,
  ``Top quarks, axigluons and charge asymmetries at hadron colliders,''
  Phys.\ Rev.\  {\bf D77 } (2008)  014003.
  [arXiv:0709.1652 [hep-ph]].

\bibitem{Chivukula:2010fk}
  R.~S.~Chivukula, E.~H.~Simmons, C.~-P.~Yuan,
  ``Axigluons cannot explain the observed top quark forward-backward asymmetry,''
  Phys.\ Rev.\  {\bf D82 } (2010)  094009.
  [arXiv:1007.0260 [hep-ph]].

\bibitem{Bai:2011ed}
  Y.~Bai, J.~L.~Hewett, J.~Kaplan, T.~G.~Rizzo,
  ``LHC Predictions from a Tevatron Anomaly in the Top Quark Forward-Backward Asymmetry,''
  [arXiv:1101.5203 [hep-ph]].

\bibitem{Ferrario:2009bz}
  P.~Ferrario, G.~Rodrigo,
  ``Constraining heavy colored resonances from top-antitop quark events,''
  Phys.\ Rev.\  {\bf D80 } (2009)  051701.
  [arXiv:0906.5541 [hep-ph]].

\bibitem{Zerwekh:2009vi}
  A.~R.~Zerwekh,
  ``Axigluon Couplings in the Presence of Extra Color-Octet Spin-One Fields,''
  Eur.\ Phys.\ J.\  {\bf C65 } (2010)  543-546.
  [arXiv:0908.3116 [hep-ph]].

\bibitem{Chivukula:2005bn}
  R.~S.~Chivukula, E.~H.~Simmons, H.~-J.~He, M.~Kurachi, M.~Tanabashi,
  ``Deconstructed Higgsless models with one-site delocalization,''
  Phys.\ Rev.\  {\bf D71 } (2005)  115001.
  [hep-ph/0502162].

\bibitem{Chivukula:2005xm}
  R.~S.~Chivukula, E.~H.~Simmons, H.~-J.~He, M.~Kurachi, M.~Tanabashi,
  ``Ideal fermion delocalization in Higgsless models,''
  Phys.\ Rev.\  {\bf D72 } (2005)  015008.
  [hep-ph/0504114].

\end{thebibliography}
\end{document}